\documentclass[12pt,letterpaper]{article}
\usepackage[T1]{fontenc}
\usepackage{amssymb,amsmath,amsfonts,dsfont,amsthm,bm}
\usepackage{float,rotfloat} 
\usepackage{mathrsfs}
\usepackage{multirow}
\usepackage{enumerate}
\usepackage{enumitem}
\usepackage{mathtools}
\usepackage{hyperref,url}
\usepackage{graphicx}
\usepackage[hang,small,bf]{caption}
\usepackage[top=1.2in, bottom=1.2in, left=1.2in, right=1.2in]{geometry}
\usepackage{longtable}
\usepackage{color,xcolor}
\usepackage[round,sort,comma]{natbib}

\usepackage[title]{appendix}
\usepackage{etoolbox}
\usepackage[english]{babel}
\usepackage[autostyle, english = american]{csquotes}
\MakeOuterQuote{"}

\patchcmd{\appendices}{\quad}{: }{}{} 

\newcommand{\ID}{\mathds{1}}
\newtheorem{thm}{Theorem}
\newtheorem{prop}{Proposition}

\newtheorem{cor}{Corollary}

\newtheorem{conj}{Conjecture}

\numberwithin{thm}{section}
\numberwithin{prop}{section}
\numberwithin{lemma}{section}
\numberwithin{cor}{section}
\numberwithin{defn}{section}
\numberwithin{conj}{section}

\numberwithin{equation}{section}
\numberwithin{figure}{section}
\numberwithin{table}{section}

\definecolor{darkblue}{rgb}{0,0,0.55}

\allowdisplaybreaks

\newcommand{\Cov}{\mathbb{C}ov}

\hypersetup{,colorlinks=true,allcolors=darkblue}

\begin{document}

\baselineskip 6mm

\thispagestyle{empty}

\begin{center}
{\large  
Robust Estimation of the Tail Index of a \\[5pt]
Single Parameter Pareto Distribution from Grouped Data
}

\vspace{15mm}

{\large\sc
Chudamani Poudyal\footnote[1]{ 
Chudamani Poudyal, PhD, ASA, 
is an Assistant Professor 
in the Department of Mathematical Sciences, 
University of Wisconsin-Milwaukee,
P.O. Box 413, 
Milwaukee, WI 53201, USA. 
~~ {\em E-mail\/}: ~{\tt cpoudyal@uwm.edu}}}

\vspace{1mm}

{\large\em University of Wisconsin-Milwaukee}

\vspace{15mm}

\copyright \
Copyright of this Manuscript is held by the Author! 
\end{center}

\vspace{3mm}

\begin{quote}
\begin{center}
{\bf\em Abstract}
\end{center}

Numerous robust estimators exist as alternatives 
to the maximum likelihood estimator (MLE) 
when a completely observed ground-up 
loss severity sample dataset is available.
However, 
the options for robust alternatives to 
MLE become significantly limited when dealing 
with grouped loss severity data,
with only a handful of methods like 
least squares, 
minimum Hellinger distance, and 
optimal bounded influence function available. 
This paper introduces a novel robust estimation technique, 
the 
{\em \underline{M}ethod of 
\underline{T}r\underline{u}ncated 
\underline{M}oments} (MTuM), 
specifically designed to estimate
the tail index of a Pareto distribution from grouped data. 
Inferential justification of MTuM is established 
by employing the central limit theorem and 
validating them through a comprehensive simulation study.

\vspace{4mm}

{\bf\em Keywords \& Phrases\/}. 
Claim Severity;
Exponential Distribution; 
Grouped Data;
Pareto Distribution;
Relative Efficiency;
Robust Estimation; 
Truncated Moments.
\end{quote}

\newpage 

\setcounter{page}{1}

\baselineskip 9.00mm

\section{Introduction}
\label{sec:Introduction}

To protect the privacy of policyholders 
(e.g., individuals, small businesses, 
privately owned companies, local government funds), 
data vendors and publicly available databases 
provide summarized data, 
that is, in a grouped format.
For statistical inference, 
we view such data as {\em i.i.d.\/} 
realizations of a random variable 
that was subjected to interval 
censoring by multiple, 
say $m$, contiguous intervals. 
The existing parametric statistical estimating 
tool for such grouped sample data
is mainly dominated by {\em maximum likelihood}. 
But {\em maximum likelihood estimator} (MLE)
typically results in sensitive fitted predictive
models if the sample data is 
coming from a contaminated
distribution, \cite{MR0120720},
or if there are heavier point masses 
assigned at some particular values,
\cite{pzb23}, 
specially for actuarial 
{\em payment-per-payment} and 
{\em payment-per-loss} data scnarios.
To overcome the sensitivity of MLE, 
several robust estimation techniques are 
established in the literature for different 
sample data scenarios except for grouped data. 
Therefore, the motivation of this scholarly 
work is to fulfil this gap by proposing a 
robust estimation method and for grouped data. 

A general class of $L$--statistics, 
\cite{MR0203874},
provides a broad selection of robust 
estimators including methods of 
{\em trimmed moments\/} (MTM) and 
{\em winsorized moments\/} (MWM).
\cite{MR2497558} and \cite{MR3758788} 
have implemented MTM and MWM, respectively,  
in actuarial framework but for completely 
observed ground-up loss severity data.
For incomplete loss data scenarios,
in a series of papers, 
\cite{MR4263275, pb22, pb23, pzb23}
consider both MTM and MWM with comprehensive 
simulation and sensitivity analysis. 
In these papers, 
it has been shown that trimming and winsorizing 
are useful methods of robustifying 
moment estimation under extreme claims, 
\cite{gw23}.
\cite{MR4192140} introduced a novel method, 
called the 
{\em \underline{M}ethod of 
\underline{T}r\underline{u}ncated 
\underline{M}oments} (MTuM), 
which employs fixed lower and upper truncation thresholds.
MTuM is designed to work for completely 
observed ground-up loss dataset.
In this approach, 
tail sample observations can be random.

For the grouped data, 
\cite{ag70} studied the problem of estimating 
the scale parameter of single parameter 
Pareto distribution via MLE and four 
variants of least squares.
As a robust alternative to MLE, 
\cite{MR2277743} considered approximate 
minimum Hellinger distance estimator
\citep{MR0448700, MR0448699}
for grouped data which can be 
asymptotically as efficient as the MLE. 
Under small model contaminations, 
\cite{MR1436121} established that 
the optimal bounded influence function estimators 
are more robust than the MLEs for grouped data.
The concept of optimal grouping,
in the sense of minimizing the loss of information, 
has been introduced by \cite{MR0909389}, 
but this approach is still under likelihood 
framework, \citet[][p. 82]{MR1994050}.
Therefore, 
the goal of this manuscript is to explore the 
robustness of the MTuM estimator specifically
for the tail index of 
{\em grouped single parameter Pareto distributions}
and to assess its performance against the corresponding MLE. 
Asymptotic distributional properties,
such as normality, consistency, 
and the asymptotic relative efficiency
in relation to the MLE, 
are established for the purpose of inferential justification. 
In addition, 
the paper strengthens its theoretical concepts 
with extensive simulation studies.
It is noteworthy that the moments, 
when subject to threshold truncation and/or censorship, 
are consistently finite, 
irrespective of the underlying true distribution.

The rest of the paper is structured as follows. 
In Section \ref{sec:mtum_for_grouped_data},
we briefly summarize the grouped
data scenarios including
different probablility functions. 
Section \ref{sec:gMTUM}
is focused on the development of 
MTuM procedures for grouped data along with
the inferential justification.
In Section \ref{sec:SimStudy},
we conduct an extensive simulation study 
to complement the theoretical results
for different scenarios.
Finally, 
concluding remarks and further directions
are summarized in Section \ref{sec:Conclusion}.

\section{Pareto Grouped Data}
\label{sec:mtum_for_grouped_data}

Due to the complexity of the involved theory, 
we only investigate single parameter Pareto
distribution in this scholarly work. 
As considered by \citet[][\S3]{MR4192140},
let $Y \sim \text{Pareto I}(\alpha,x_{0})$ 
with the distribution function 
$F_{Y}(y) = 1-\left(x_{0}/y\right)^{\alpha}, \ y > x_{0}$, 
zero elsewhere. 
Here, $\alpha > 0$ represents the shape parameter,
often referred to as the tail index, 
and $x_{0} > 0$ is the known lower bound threshold.
Consequently, 
if we define $X := log{\left(Y/x_{0}\right)}$,
then $X$ follows an exponential distribution,
$X \sim \text{Exp}(\theta = 1/\alpha)$, 
with its distribution function given by
$F_{X}(x) = 1-e^{-x/\theta}$.
Hence, estimating $\alpha$ is 
equivalent to estimating the exponential 
parameter $\theta$.
Thus, for the purpose of analytic simplicity, 
we investigate $\theta$, rather than $\alpha$. 
The development and asymptotic behavior of MTuM 
estimators will be explored for a 
grouped sample drawn from 
an exponential distribution. 

Let 
$0 < c_{1} < \cdots < c_{m-1} < c_{m} < \infty$
be the group boundaries for the grouped data 
where we define 
$c_{0} := 0$, and 
$c_{m+1} := \infty$.  
Let 
$X_{1},\ldots,X_{n} \stackrel{i.i.d.}
\sim X$,
where $X$ has pdf 
$f(x|\theta) 
= f(x) 
= \frac{1}{\theta}e^{-\frac{x}{\theta}}$ 
and cdf  
$F(x|\theta) = F(x)$.
Computation of the empirical distribution function at the group boundaries is clear, 
but inside the intervals, the linearly interpolated empirical cdf as defined in 
\citet[][\S14.2]{MR3890025},  
is the most common one.
The linearly interpolated empirical cdf, called "ogive" and denoted by $F_{n}$, 
is defined as
\begin{equation} 
\label{eq:ogive}
F_{n}(x)
=
\begin{cases}
\frac{c_{j}-x}{c_{j}
-
c_{j-1}}F_{n}(c_{j-1})
+
\frac{x-c_{j-1}}{c_{j}-c_{j-1}}F_{n}(c_{j}); & 
\mbox{if }
c_{j-1} < x \leq c_{j}, \ \ j \leq m, \\ 
\mbox{Undefined}; & 
\mbox{if } x > c_{m}.
\end{cases}
\end{equation}
In the {\em complete\/} data case, we observe the following 
{\em empirical\/} frequencies of $X$:
\[
\widehat{\mathbb{P}} \big[ c_{j-1} < X \leq c_j \big] = 
F_n (c_j) - F_n (c_{j-1}) =
\frac{n_j}{n}, \qquad j = 1, \ldots, m+1,
\]
where 
$
\displaystyle 
n_{j}
= 
\sum_{i=1}^{n}\ID\{ c_{j-1} < X_{i} \le c_{j}\},
$
giving 
$
\displaystyle 
n 
=
\sum_{j=1}^{m+1} n_j
$ 
is the sample size.

Clearly, 
the empirical distribution $F_{n}$ is 
not defined in the interval 
$(c_{m},c_{m+1}=\infty)$ as it 
is impossible to draw a straight 
line joining two points
$\left(c_{m},F_{n}(c_{m})\right)$ 
and $(\infty,1)$
unless $F_{n}(c_{m}) = 1$. 

The corresponding linearized population cdf 
$F_{\mbox{\tiny G}}$ 
is defined by
\begin{equation} 
\label{eq:pop_ogive}
F_{\mbox{\tiny G}}(x)=
\left\{
\begin{array}{ll}
\frac{c_{j}-x}{c_{j}-c_{j-1}}
F(c_{j-1}|\theta)+\frac{x-c_{j-1}}{c_{j}-c_{j-1}}F(c_{j}|\theta); &  
\mbox{if } 
c_{j-1} < x \leq c_{j}, \ \ j \leq m, \\[2ex]
F(x|\theta); & 
\mbox{if } x > c_{m}. \\
\end{array}
\right.
\end{equation}

The corresponding density function $f_{n}$,
called the histogram, is defined as
\begin{equation} 
\label{eq:histGroupedData}
f_{n}(x)
=
\begin{cases}
\frac{F_{n}(c_{j})-F_{n}(c_{j-1})}{c_{j}-c_{j-1}}=\frac{n_{j}}{n(c_{j}-c_{j-1})}; &
\mbox{if }
c_{j-1} < x \leq c_{j}, \ \ j \leq m, \\
\mbox{Undefined}; & 
\mbox{if } x > c_{m}.
\end{cases} 
\end{equation}
The empirical quantile function 
(the inverse of $F_{n}$)
is then computed as
\begin{equation} 
\label{eq:inv_ogive}
F_{n}^{-1}(s)
=
\begin{cases}
c_{j-1}
+
\frac{(c_{j}
-
c_{j-1})(s-F_{n}(c_{j-1}))}{F_{n}(c_{j})
-
F_{n}(c_{j-1})}; & 
\mbox{if }
F_{n}(c_{j-1}) < s 
\leq 
F_{n}(c_{j}), \ j \leq m, \\ 
\mbox{Undefined}; & 
\mbox{if } s > F_{n}(c_{m}).
\end{cases}
\end{equation}
Similarly,
\begin{equation} 
\label{eq:inv_pop_ogive}
F_{\mbox{\tiny G}}^{-1}(s|\theta)
=
\left\{
\begin{array}{ll}
c_{j-1}+\frac{(c_{j}-c_{j-1})(s-F(c_{j-1}|\theta))}{F(c_{j}|\theta)-F(c_{j-1}|\theta)},  
&  F(c_{j-1}|\theta) < s \leq F(c_{j}|\theta), \ j \leq m; \\[2ex]
F^{-1}(s|\theta), & s > F(c_{m}|\theta). \\
\end{array}
\right.
\end{equation}

If the loss variable $X$ observed in a grouped format 
is affected by additional transformations: 
truncation, interval censoring, 
coverage modifications,
then in those cases, the underlying distribution 
function would have to be modified accordingly. 
For example, if $m$ 
groups ($n$ observations in total) are provided and it is known 
that only data above deductible $d$ appeared,
then the distributional assumption is that we observe
\[
\widehat{\mathbb{P}} 
\left[ c_{j-1} < X \leq c_j \, \big | \, X > d \right] = 
\frac{n_j}{n}, \qquad j = 1, \ldots, m+1,
\]
with the group boundaries satisfying 
$d = c_0 < c_1 < \cdots < c_{m} < c_{m+1} = \infty$.

\section{MTuM for Grouped Data}
\label{sec:gMTUM}

For both MTM and MWM,
if the right trimming/winsorzing proportion
$b$ is such that $1 - b > F_{n}(c_{m})$,
then we have 
$c_{m} < F_{n}^{-1}(1-b) < c_{m+1} = \infty$.
That is, $F_{n}^{-1}(1-b)$ does not exist
as the linearized empirical distribution 
$F_{n}$ is not defined in the interval 
$(c_{m},c_{m+1}=\infty)$, 
see Eq. \eqref{eq:ogive}.
As a consequence, 
$F_{n}^{-1}$ is not defined on the 
interval $(F_{n}(c_{m}),1]$.
Thus, 
in order to apply the MTM/MWM
approach for grouped sample,
we always need to make sure that
$F_{n}^{-1}(1-b) \leq c_{m}$, 
that is, $1-b \leq F_{n}(c_{m})$,
but this is problematic for different
samples with the fixed right 
trimming/winsorzing $b$.
With this fact in consideration, 
the asymptotic distributional 
properties of MTM and MWM 
estimators and from grouped 
data are very complicated
and not easy to analytically 
derive if not intractable.
But with MTuM, 
we can always choose the right
truncated threshold $T$ such that 
$T \le c_{m}$.
Therefore, 
we proceed with MTuM approach 
for grouped data in the rest 
of this section. 
Let $0 \le t$ and $T \le c_{m}$,
with $t < T$,
be the left and right truncation points,
respectively. 

By using the empirical cdf, Eq. (\ref{eq:ogive}) and 
pdf Eq. (\ref{eq:histGroupedData}), 
the sample truncated moments 
for a grouped data as defined by 
\cite{MR4192140} is given by

\begin{equation}
\widehat{\mu} 
=
\frac{1}{F_{n}(T)
-F_{n}(t)}
{\int_{t}^{T}{h(x)f_{n}(x) \, dx}}.
\label{eq:sample_mtum_group}
\end{equation}
Let us introduce the following notations:
\begin{align*}
\left.
\begin{array}{lrl}
p_{j} 
& 
\equiv 
p_{j}(\theta) 
& :=
F(c_{j}|\theta)  \\ [10pt] 
P_{j} 
& 
\equiv
P_{j}(\theta) 
& :=
F(c_{j}|\theta)-F(c_{j-1}|\theta) \\[10pt]
{} & 
p_{j,n} 
& :=
F_{n}(c_{j}) \\ [10pt] 
{} & 
\sigma_{j,j'}^{2} 
& := 
\Cov
\left(
F_{n}(c_{j}),F_{n}(c_{j'})
\right) \\ [10pt]
{} & {}
& = 
\Cov
\left(p_{j},p_{j'}\right) \\ [10pt]
{} & 
I_{i,j} 
& :=
\ID\{X_{i} \leq c_{j}\} \\ [10pt]
{} & 
J_{i,j} 
& :=
\ID\{X_{i} > c_{j}\}
\end{array} 
\qquad \right\} \text{  for  } 0 \leq j,j' \leq m+1; 0 \leq i \leq n. 
\end{align*}

\begin{prop} 
\label{prop:piqj}
Suppose $1 \leq j\leq j' \leq m$. 
Then 
$
\Cov
\left(
p_{j,n},1-p_{j',n}
\right) 
= 
-\frac{p_{j}(1-p_{j'})}{n}$.
\begin{proof}
Clearly, 
$
p_{j,n} = (1/n)\sum_{i=1}^{n}I_{i,j}
$
and 
$
1 - p_{j',n} 
= 
(1/n)\sum_{i=1}^{n}J_{i,j'}.
$
Therefore,
\begin{align*}
\Cov
\left(
p_{j,n},1-p_{j',n}
\right) 
& = 
\Cov
\left(
\frac{1}{n}\sum_{i=1}^{n}I_{i,j},
\frac{1}{n}\sum_{i=1}^{n}J_{i,j'}
\right) 
= 
\frac{1}{n^{2}}\Cov
\left(\sum_{i=1}^{n}I_{i,j}, \sum_{i=1}^{n}J_{i,j'}\right) \\
& =
\frac{1}{n^{2}}
\sum_{k=1}^{n}
\sum_{i=1}^{n}
\Cov\left(I_{k,j},J_{i,j'}\right) 
= 
\frac{1}{n^{2}}
\sum_{i=1}^{n}\Cov
\left(I_{i,j},J_{i,j'}\right) \\
& = 
\frac{1}{n^{2}}n\Cov
\left(I_{1,j},J_{1,j'}\right) 
= 
\frac{1}{n}\left[\mathbb{E}(I_{1,j}J_{1,j'}) 
- 
\mathbb{E}(I_{1,j})\mathbb{E}(J_{1,j'})\right] \\
& = 
\frac{1}{n}
\left[0 - p_{j}(1-p_{j'})\right] 
= 
-\frac{p_{j}(1-p_{j'})}{n}. \qedhere
\end{align*}
\end{proof}
\end{prop}
The following corollary is an immediate consequence of Proposition \ref{prop:piqj}. 
\begin{cor}
Let $(F_{n}(c_{1}),\ldots,F_{n}(c_{m}))$ be a vector of empirical distribution function evaluated at the group boundaries vector $(c_{1},\ldots,c_{m})$. 
Then, $(F_{n}(c_{1}), \allowbreak \ldots,F_{n}(c_{m}))$ is $\mathcal{AN}\left(\bm{F},n^{-1}\bm{\Sigma}\right)$, where $\bm{F} = (F(c_{1}|\theta),\ldots,F(c_{m}|\theta))$, $\bm{\Sigma} = \left[\sigma_{jj'}^{2}\right]_{j,j'=1}^{m}$, with $\sigma_{jj'}^{2} = \sigma_{j'j}^{2} = F(c_{j}|\theta)(1-F(c_{j'}|\theta))$ for all $j \leq j'$. 
\end{cor}

Assume that
$c_{0} 
\leq
c_{l-1} < t 
\leq
c_{l} \leq c_{r} < T 
\leq
c_{r+1} \leq c_{m}$.
Then, 
\begin{align*}
F_{n}(t) 
& =
A_{1}F_{n}(c_{l-1}) + B_{1}F_{n}(c_{l})
\quad \mbox{and} \quad 
F_{n}(T) 
=
A_{2}F_{n}(c_{r}) + B_{2}F_{n}(c_{r+1}),
\end{align*}
where 
\[
A_{1} 
:=
\frac{c_{l}-t}{c_{l}-c_{l-1}}, \quad 
A_{2} 
:=
\frac{c_{r+1}-T}{c_{r+1}-c_{r}}, \quad
B_{1} 
:=
\frac{t-c_{l-1}}{c_{l}-c_{l-1}}, 
\quad \mbox{and} \quad 
B_{2}
:=
\frac{T-c_{r}}{c_{r+1}-c_{r}}.
\] 
Also, consider 
\[
u_{l} 
:=
\frac{c_{l}^{2}-t^{2}}
{2(c_{l} - c_{l-1})}, \quad
v_{i} 
:=
\frac{c_{i}+c_{i-1}}{2},
\quad \mbox{and} \quad 
z_{r} 
:=
\frac{T^{2}-c_{r}^{2}}{2(c_{r+1}-c_{r})}.
\]
Assuming $h(x) \equiv x$,
after some computation,
we get
\begin{align*}
g_{\mu}(p_{1,n},\ldots,p_{m,n}) & := \widehat{\mu} \\ 
& = \frac{u_{l}(p_{l,n}-p_{l-1,n})+\sum_{i=l+1}^{r}v_{i}(p_{i,n}-p_{i-1,n})+z_{r}(p_{r+1,n}-p_{r,n})}{A_{2}p_{r,n}+B_{2}p_{r+1,n}-A_{1}p_{l-1,n}-B_{1}p_{l,n}} 
=: 
\frac{N}{H}.
\end{align*}

Note that $p_{0,n}, = 0$. Thus, by the delta method 
\cite[see, e.g.,][Theorem A, p. 122]{MR595165},
we have
\[
\widehat{\mu} 
\sim
\mathcal{AN}\left(\mu = g_{\mu}
(\bm{F}),n^{-1}\bm{D}_{\mu}\bm{\Sigma} \bm{D}_{\mu}'\right),
\]
where $\bm{D}_{\mu} := \left(\left(\frac{\partial g_{\mu}}{\partial p_{1,n}},\ldots,\frac{\partial g_{\mu}}{\partial p_{m,n} } \right)_{\bm{p} = \bm{F}}\right)$ and $\bm{p} := \left(p_{1,n},\ldots, p_{m,n}\right)^{'}$. Consider $\bm{\Sigma}_{\mu} := \bm{D}_{\mu}\bm{\Sigma} \bm{D}_{\mu}'$. 
Clearly, if $2 \leq l<r$ then
\begin{align*}
\frac{\partial g_{\mu}}{\partial p_{j,n}} & = 
\begin{cases}
0, & \ \ \ \ \ \text{for } 1 \leq j \leq l-2 \text{ or } j \geq r+2; \\[10pt]
\frac{-u_{l}H+A_{1}N}{H^{2}}, & \ \ \ \ \ \text{for } j = l-1; \\[10pt]
\frac{(u_{l}-v_{l+1})H+B_{1}N}{H^{2}}, & \ \ \ \ \ \text{for } j = l; \\[10pt]
\frac{c_{j-1} - c_{j+1}}{2H}, & \ \ \ \ \ \text{for } l+1 \leq j \leq r-1; \\[10pt]
\frac{(v_{r}-z_{r})H-A_{2}N}{H^{2}}, & \ \ \ \ \ \text{for } j = r; \\[10pt]
\frac{z_{r}H-B_{2}N}{H^{2}}, & \ \ \ \ \ \text{for } j = r+1. \\[10pt]
\end{cases}
\end{align*}
And if $l=r$, 
\begin{align*}
\frac{\partial g_{\mu}}{\partial p_{j,n}} & = 
\begin{cases}
0, & \ \ \ \ \ \text{for } 1 \leq j \leq l-2 \text{ or } j \geq l+2; \\[10pt]
\frac{-u_{l}H+A_{1}N}{H^{2}}, & \ \ \ \ \ \text{for } j = l-1; \\[10pt]
\frac{(u_{l}-z_{r})H-(A_{2}-B_{1})N}{H^{2}}, & \ \ \ \ \ \text{for } j = l; \\[10pt]
\frac{z_{r}H-B_{2}N}{H^{2}}, & \ \ \ \ \ \text{for } j = l+1. \\[10pt]
\end{cases}
\end{align*}

\begin{figure}[hbt!]
\centering
\includegraphics[width=0.48\textwidth]
{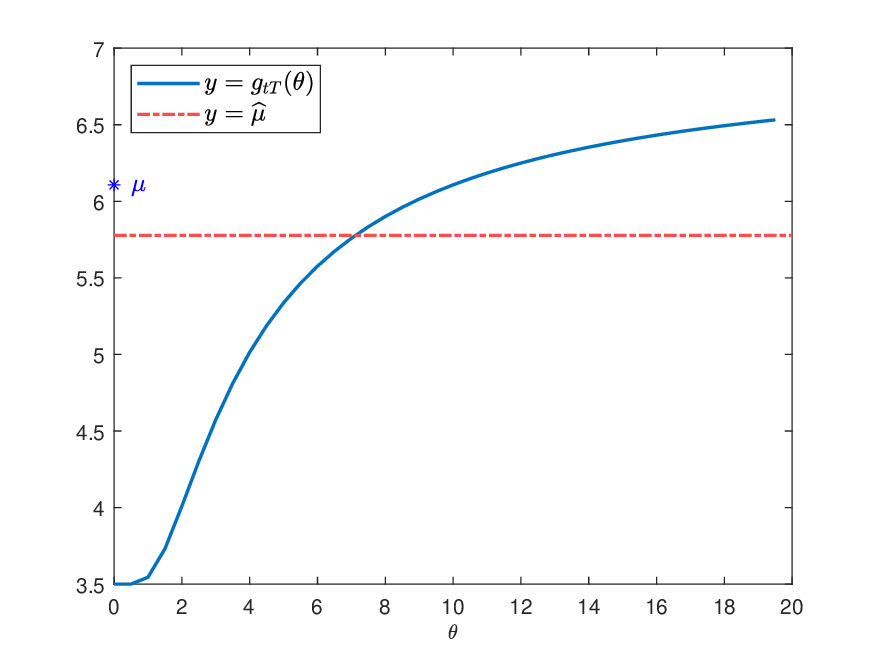}
\includegraphics[width=0.48\textwidth]
{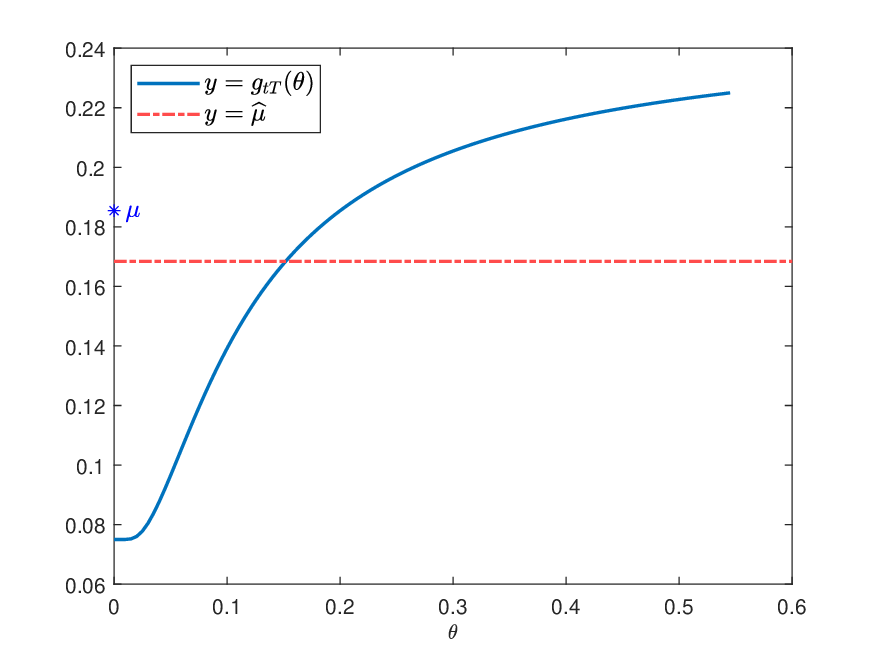}
\caption
{
Graphs of $g_{tT}$ for different values of $\theta$.
Left panel represents the graph of $g_{tT}(\theta)$
for $\theta = 10$, $(t,T) = (2,12)$, and
group boundaries vector
$v_{1} = (0,5,10,15,20,25)$.
Similarly, right panel represents the graph of $g_{tT}(\theta)$
for $\theta = 0.2$, $(t,T) = (.05,.45)$, and
group boundaries vector
$v_{2} = (0,.1,.2,.3,.4,.5)$.
}
\label{fig:yFigDec_20_2020}
\end{figure}

By using Eq. \eqref{eq:pop_ogive}, 
the corresponding linearized 
population mean  is
\begin{align} 
\label{eqn:gTt1} 
g_{tT}(\theta)
& : =
\mu
= 
\frac{u_{l}P_{l}(\theta)+\sum_{i=l+1}^{r}v_{i}P_{i}(\theta)+z_{r}P_{r+1}(\theta)}{A_{2}p_{r}(\theta)+B_{2}p_{r+1}(\theta)-A_{1}p_{l-1}(\theta)-B_{1}p_{l}(\theta)}
= \frac{N^{*}}{H^{*}}.
\end{align} 
Due to the intense nature of the function $g_{tT}(\theta)$, 
it is complicated to come up with an analytic justification
establishing whether it is increasing or decreasing. 
But at least for $X \sim \mbox{Exp}(\theta)$, 
$g_{tT}(\theta)$ appears to be an increasing 
function of $\theta > 0$ as shown in 
Figure \ref{fig:yFigDec_20_2020}.
Generally, we summarize the result 
in the following conjecture.

\begin{conj} 
\label{conj:gtTGrouped}
The function $g_{tT}(\theta)$ 
is strictly increasing. 
\end{conj}

\begin{prop}
\label{prop:gtTLimGroupedExp}
The function $g_{tT}(\theta)$
has the following limiting values
\begin{align}
\lim_{\theta \rightarrow 0+}g_{tT}(\theta) 
& = 
\frac{u_{l}}{A_{1}}, \\[2ex]
\lim_{\theta \rightarrow \infty}g_{tT}(\theta) 
& = 
\frac{u_{l}(c_{l-1}-c_{l})
+
\sum_{i=l+1}^{r}v_{i}(c_{i-1}-c_{i})
+
z_{r}(c_{r}-c_{r+1})}
{A_{1}c_{l-1}+B_{1}c_{l}
-A_{2}c_{r}
-
B_{2}c_{r+1}}.
\end{align}
\begin{proof}
These limits can be established by using 
L'H{\^o}pital's rule.
\end{proof}
\end{prop}

\noindent
Now, assuming the Conjecture \ref{conj:gtTGrouped} 
is true then with Proposition \ref{prop:gtTLimGroupedExp}, 
we have 

\begin{thm}
The equation $\widehat{\mu} 
=
g_{tT}(\theta)$ 
has a unique solution 
$\widehat{\theta}_{\mbox{\tiny MTuM}}$ 
provided that 
\[
\frac{u_{l}}{A_{1}} 
<
\widehat{\mu} 
<
\frac{u_{l}(c_{l-1}-c_{l})
+ 
\sum_{i=l+1}^{r}v_{i}(c_{i-1}-c_{i})
+
z_{r}(c_{r}-c_{r+1})}
{A_{1}c_{l-1}
+
B_{1}c_{l}
-A_{2}c_{r}-B_{2}c_{r+1}}.
\]
\end{thm}

Solve the equation $\widehat{\mu} = \mu$ for 
$\widehat{\theta}_{\mbox{\tiny MTuM}}$,
say 
$\widehat{\theta} 
=: 
g_{\theta}(\widehat{\mu})$. 
Then, again by the delta method, we conclude that $\widehat{\theta}\sim \mathcal{AN}\left(g_{\theta}(\mu), n^{-1}\left(g_{\theta}'(\mu)\right)^{2}\bm{\Sigma}_{\mu}\right)$. 
Note that if both the left- and right-truncation points lie on the same interval, then $\widehat{\mu} = \frac{t+T}{2} = \mu$. 
So, the parameter to be estimated disappears from the 
equation and hence we do not consider this case for further investigation.
Define
\begin{align*}
P 
& :=
u_{l}
\left(
e^{-\frac{c_{l-1}}{\theta}}-e^{-\frac{c_{l}}{\theta}}
\right) 
+
\sum_{i=l+1}^{r}v_{i}
\left(
e^{-\frac{c_{i-1}}{\theta}}-e^{-\frac{c_{i}}{\theta}}
\right)
+z_{r}
\left(
e^{-\frac{c_{r}}{\theta}}-e^{-\frac{c_{r+1}}{\theta}}
\right), \\
Q 
& :=
B_{2}
\left(
1-e^{-\frac{c_{r+1}}{\theta}}
\right)
-
A_{1}
\left(
1-e^{-\frac{c_{l-1}}{\theta}}
\right)
-
B_{1}
\left(
1-e^{-\frac{c_{l}}{\theta}}
\right).
\end{align*}
Then, we get a fixed point function as $\theta = G(\theta)$, 
where 
\begin{align}
G(\theta) 
& =
-\frac{c_{r}}{\log\left(\frac{\widehat{\mu}A_{2}-P+\widehat{\mu}Q}{\widehat{\mu}A_{2}}\right)}.
\end{align}
However, we need to consider the condition $\widehat{\mu}(A_{2}+Q)>P$. Therefore, we need to be careful about the initialization of $\theta$
as the right truncation point $T$ cannot be a boundary point. 
Because if it was, then $A_{2}=0$ and we would not be able to divide by
$A_{2}$ in the fixed point function $\theta = G(\theta)$.

Now, let us compute the derivative of $g_{\theta}$ with respect to $\mu$, using implicit differentiation.  

\noindent
\textbf{Case 1}: Assume that the two truncation points are in two consecutive intervals, i.e., assume that $l=r$. Then $\theta' = g'_{\theta}(\widehat{\mu}) = \frac{A-B}{\Lambda + \Delta}$, where
\begin{align*}
A 
& :=
A_{2}+B_{2}-A_{1}-B_{1}, 
\quad 
B 
:= 
A_{2}e^{-\frac{c_{r}}{\theta}} + B_{2}e^{-\frac{c_{r+1}}{\theta}} - A_{1}e^{-\frac{c_{l-1}}{\theta}} - B_{1}e^{-\frac{c_{l}}{\theta}}, \\
\Lambda & := \frac{u_{l}}{\theta^{2}}\left(c_{l-1}e^{-\frac{c_{l-1}}{\theta}}-c_{l}e^{-\frac{c_{l}}{\theta}}\right) + \frac{z_{r}}{\theta^{2}}\left(c_{r}e^{-\frac{c_{r}}{\theta}}-c_{r+1}e^{-\frac{c_{r+1}}{\theta}}\right), \\
\Delta & := \frac{\widehat{\mu}}{\theta^{2}}\left(A_{2}c_{r}e^{-\frac{c_{r}}{\theta}} + B_{2}c_{r+1}e^{-\frac{c_{r+1}}{\theta}} - A_{1}c_{l-1}e^{-\frac{c_{l-1}}{\theta}} - B_{1}c_{l}e^{-\frac{c_{l}}{\theta}}\right).
\end{align*}
\textbf{Case 2}:
The other case is that the two truncation points are not 
in the two consecutive intervals, i.e., assume that $l<r$. 
Then 
$\theta'
=
g'_{\theta}(\widehat{\mu}) 
=
\frac{A-B}{\Gamma + \Delta}$, 
where 
$
\displaystyle 
\Gamma 
:=
\Lambda + \sum_{i=l+1}^{r} 
\frac{v_{i}}{\theta^{2}}
\left(c_{i-1}
e^{-\frac{c_{i-1}}{\theta}}
-c_{i}e^{-\frac{c_{i}}{\theta}}\right)
$ 
and $A, \ B, \ \Lambda$, and $\Delta$ are defined above.

To get exponential grouped MLE, 
consider 
$
\displaystyle 
P_{j}(\theta) 
:= 
e^{-\frac{c_{j-1}}{\theta}} - e^{-\frac{c_{j}}{\theta}}.
$ 
Then, following \cite{MR1900948},
we have, 
$\widehat{\theta}_{\mbox{\tiny MLE}} 
\sim
\mathcal{AN}\left(\theta, 
\frac{1}{n} \bm{I}^{-1}(\theta)\right)$, 
where 
$
\displaystyle 
\bm{I}(\theta) = \sum_{j=1}^{m}P_{j}(\theta)
\left(\frac{d\ln{P_{j}(\theta)}}{d\theta}\right)^{2}$. 
Note that after finding the derivative, $\bm{I}(\theta)$ 
can be expressed as 
\begin{align*}
\bm{I}(\theta) 
& =
\sum_{j=1}^{m}{P_{j}(\theta) 
\left(\frac{c_{j-1}e^{-\frac{c_{j-1}}{\theta}} 
- c_{j}e^{-\frac{c_{j}}{\theta}}}
{\theta^2 
\left(
e^{-\frac{c_{j-1}}{\theta}} 
- e^{-\frac{c_{j}}{\theta}}
\right)}
\right)^{2}} 
= 
\sum_{j=1}^{m}
\left(\frac{c_{j-1}e^{-\frac{c_{j-1}}{\theta}} 
- c_{j}e^{-\frac{c_{j}}{\theta}}}{\theta^2}\right)^{2}
\frac{1}{P_{j}(\theta)}.
\end{align*}

The asymptotic performance of MTuM estimator 
is measured through the asymptotic relative efficiency (ARE) 
in comparison to the grouped MLE. 
ARE citep[see, e.g.,][]{MR595165,MR1652247} is defined as:
\begin{equation} 
\label{eqn:ARE1}
ARE(MTuM, MLE) 
=
\frac{\text{asymptotic variance of MLE estimator}}
{\text{asymptotic variance of MTuM estimator}}.
\end{equation}
The primary justification for employing MLE as a standard/benchmark 
for comparison lies in its optimal asymptotic behavior in terms of variance, 
though this comes with the typical proviso of being subject to '
"under certain regularity conditions".
Therefore, the desired ARE as given by Eq. \eqref{eqn:ARE1} is computed as 
\begin{align}
ARE
\left(
\widehat{\theta}_{\mbox{\tiny MTuM}}, \widehat{\theta}_{\mbox{\tiny MLE}}
\right)
& =
\frac{\bm{I}^{-1}(\theta)}{(g_{\theta}'
({\mu}))^{2}
{\bm{\Sigma}}_{\mu}} 
= 
\frac{\bm{I}^{-1}(\theta)}{(g_{\theta}'
({\mu}))^{2}\bm{D}_{\mu}\bm{\Sigma} \bm{D}_{\mu}'}.
\end{align}
The numerical values of 
$
ARE
\left(
\widehat{\theta}_{\mbox{\tiny MTuM}}, \widehat{\theta}_{\mbox{\tiny MLE}}
\right)
$
from $\mbox{Exp}(\theta = 10)$
with group boundaries vector
$G := (0 : 4 : 30,  \infty)$
is summarized in  
Table \ref{table:gARE1}.

\begin{table}[hbt!] 
\centering
\begin{tabular}{c|ccccc}
\hline \\[-2.25ex]
\multicolumn{1}{c|}{} &		
\multicolumn{5}{c}{T{\tiny $(1-F(T))$}} \\
t{\tiny (F(t))} & $30${\tiny (.05)} & 23{\tiny (.10)} & 19{\tiny (.15)} & 14{\tiny (.25)} & 7{\tiny (.5)} \\
\hline\hline
& & & & & \\[-2.25ex]
00.0{\tiny (.00)} & 
.493 & .346 & .234 & .121 & .040 \\
00.5{\tiny (.05)} &
.492 & .347 & .235 & .122 & .040 \\
01.0{\tiny (.10)} & 
.489 & .348 & .235 & .123 & .040 \\
01.5{\tiny (.14)} & 
.483 & .346 & .234 & .123 & .040 \\
03.0{\tiny (.26)} & 
.429 & .313 & .210 & .115 & .040 \\
07.0{\tiny (.50)} & 
.212 & .136 & .074 & .024 & - \\
14.0{\tiny (.75)} & 
.057 & .037 & .015 & - & - \\
19.0{\tiny (.85)} & 
.017 & .009 & - & - & - \\
23.0{\tiny (.90)} & 
.005 & - & - & - & - \\
\hline\hline
\end{tabular}
\caption{$ARE\left(\widehat{\theta}_{\mbox{\tiny MTuM}},\widehat{\theta}_{MLE}\right)$ 
for selected $t$ and $T$ 
with $G = (0:5:30,\infty)$
from $\mbox{Exp}(\theta = 10)$.}
\label{table:gARE1}
\end{table}

\section{Simulation Study} 
\label{sec:SimStudy}

This section augments the theoretical findings 
established in Section \ref{sec:gMTUM} with simulations.
The primary objective is to determine the sample size
required for the estimators to be unbiased 
(acknowledging that they are asymptotically unbiased), 
to validate the asymptotic normality, 
and to ensure that their finite 
sample relative efficiencies (RE) 
are converging towards the respective AREs.
For calculating RE, 
MLE is utilized as the reference point.
Consequently, 
the concept of asymptotic relative efficiency 
outlined in equation (\ref{eqn:ARE1}) 
is adapted for finite sample analysis as follows:
\begin{equation*} 
\label{eq:finite_relative_efficiency_benchmark_MLE}
RE(MTuM, MLE) 
=
\frac{\text{asymptotic variance of MLE estimator}}
{\text{variance of a competing estimator MTuM}}.
\end{equation*}

\begin{table}[hbt!] 
\centering
\begin{tabular}{ccc|cccccc|cc}
\hline\hline
\multicolumn{1}{c}{} &
\multicolumn{2}{c|}{} &
\multicolumn{8}{|c}{{MTuM Performance for Exponential Grouped Data}} \\
\cline{4-11}
\multicolumn{1}{c}{} &
\multicolumn{2}{c|}{} &
\multicolumn{8}{|c}{$n$} \\
\cline{4-11}
& $t_{l}$ & $t_{r}$ & 50 & 100 & 250 & 500 & 1000 & $\infty$ & $\infty$ & $\infty$ \\
\hline
& & & & & & & & & & \\[-2.25ex]
MEAN & 0 & 200 &
1.00{\tiny (.003)} & 1.00{\tiny (.003)} & 1.00{\tiny (.002)} &
1.00{\tiny (.001)} & 1.00{\tiny (.001)} & 1 & - & - \\
& 0 & 50 &
1.01{\tiny (.004)} & 1.00{\tiny (.002)} & 1.00{\tiny (.002)} &
1.00{\tiny (.001)} & 1.00{\tiny (.001)} & 1 & - & - \\
& 0 & 100 &
1.00{\tiny (.003)} & 1.00{\tiny (.003)} & 1.00{\tiny (.001)} &
1.00{\tiny (.001)} & 1.00{\tiny (.001)} & 1 & - & - \\
& 0 & 140 &
1.00{\tiny (.005)} & 1.00{\tiny (.004)} & 1.00{\tiny (.002)} &
1.00{\tiny (.001)} & 1.00{\tiny (.001)} & 1 & - & - \\
& 2 & 12 &
3.22{\tiny (.174)} & 1.68{\tiny (.130)} & 1.14{\tiny (.021)} &
1.06{\tiny (.012)} & 1.03{\tiny (.005)} & 1 & - & - \\
\hline\hline
& & & & & & & & & & \\[-2.25ex]
{\sc RE} & 0 & 200 &
1.01{\tiny (.032)} & 1.02{\tiny (.037)} & 1.02{\tiny (.039)} &
0.99{\tiny (.045)} & 0.99{\tiny (.048)} & 1.00 & 1.00 & 1.00 \\
& 0 & 50 &
0.77{\tiny (.048)} & 0.79{\tiny (.058)} & 0.81{\tiny (.032)} &
0.84{\tiny (.035)} & 0.83{\tiny (.028)} & 0.82 & 0.82 & 1.00 \\
& 0 & 100 &
1.00{\tiny (.045)} & 0.98{\tiny (.040)} & 1.02{\tiny (.050)} &
0.99{\tiny (.034)} & 0.99{\tiny (.046)} & 1.00 & 0.99 & 1.00 \\
& 0 & 140 & 
0.96{\tiny (.042)} & 1.01{\tiny (.063)} & 0.97{\tiny (.047} &
1.00{\tiny (.047)} & 1.01{\tiny (.033)} & 1.00 & 1.00 & 1.00 \\
& 2 & 12 &
0.00{\tiny (.000)} & 0.00{\tiny (.000)} & 0.01{\tiny (.004)} &
0.02{\tiny (.004)} & 0.03{\tiny (.002)} & 0.04 & 0.04 & 1.00 \\
\hline\hline 
\end{tabular} 
\caption{Finite-sample performance evaluation of MTuM w.r.t. MLE for grouped data from $Exp(\theta=10)$ with group boundaries vector $G_{1} = \left(0:1:100,200\right)$.}
\label{table:MTuM_sim_grouped_exp1}
\end{table}

\begin{table}[hbt!] 
\centering
\begin{tabular}{ccc|cccccc|cc}
\hline\hline
\multicolumn{1}{c}{} &
\multicolumn{2}{c|}{} &
\multicolumn{8}{|c}{{MTuM Performance for Exponential Grouped Data}} \\
\cline{4-11}
\multicolumn{1}{c}{} &
\multicolumn{2}{c|}{} &
\multicolumn{8}{|c}{$n$} \\
\cline{4-11}
& $t_{l}$ & $t_{r}$ & 50 & 100 & 250 & 500 & 1000 & $\infty$ & $\infty$ & $\infty$ \\
\hline
& & & & & & & & & & \\[-2.25ex]
MEAN & 0 & 200 &
1.00{\tiny (.006)} & 1.00{\tiny (.003)} & 1.00{\tiny (.002)} &
1.00{\tiny (.001)} & 1.00{\tiny (.001)} & 1 & - & - \\
& 0 & 50 &
1.01{\tiny (.005)} & 1.00{\tiny (.003)} & 1.00{\tiny (.002)} &
1.00{\tiny (.002)} & 1.00{\tiny (.001)} & 1 & - & - \\
& 0 & 100 &
1.00{\tiny (.006)} & 1.00{\tiny (.003)} & 1.00{\tiny (.002)} &
1.00{\tiny (.002)} & 1.00{\tiny (.001)} & 1 & - & - \\
& 0 & 140 &
1.00{\tiny (.005)} & 1.00{\tiny (.004)} & 1.00{\tiny (.002)} &
1.00{\tiny (.001)} & 1.00{\tiny (.001)} & 1 & - & - \\
& 2 & 12 &
3.17{\tiny (.169)} & 1.67{\tiny (.083)} & 1.16{\tiny (.024)} &
1.05{\tiny (.007)} & 1.03{\tiny (.006)} & 1 & - & - \\
\hline\hline
& & & & & & & & & & \\[-2.25ex]
{\sc RE} & 0 & 200 &
1.00{\tiny (.061)} & 0.99{\tiny (.071)} & 0.99{\tiny (.047)} &
0.98{\tiny (.068)} & 1.04{\tiny (.049)} & 1.00 & 1.00 & 1.00 \\
& 0 & 50 &
0.75{\tiny (.054)} & 0.81{\tiny (.037)} & 0.81{\tiny (.023)} &
0.82{\tiny (.038)} & 0.84{\tiny (.038)} & 0.82 & 0.82 & 1.00 \\
& 0 & 100 &
0.99{\tiny (.047)} & 0.96{\tiny (.039)} & 0.99{\tiny (.065)} &
1.03{\tiny (.043)} & 0.99{\tiny (.057)} & 1.00 & 0.99 & 1.00 \\
& 0 & 140 & 
0.99{\tiny (.028)} & 1.02{\tiny (.046)} & 1.02{\tiny (.050)} &
1.00{\tiny (.041)} & 1.01{\tiny (.043)} & 1.00 & 1.00 & 1.00 \\
& 2 & 12 &
0.00{\tiny (.000)} & 0.00{\tiny (.000)} & 0.01{\tiny (.003)} &
0.02{\tiny (.004)} & 0.03{\tiny (.001)} & 0.04 & 0.04 & 1.00 \\
\hline\hline 
\end{tabular} 
\caption{Finite-sample performance evaluation of MTuM w.r.t. MLE for grouped data from $Exp(\theta=10)$ with group boundaries vector $G_{2} = 0:1:200$.}
\label{table:MTuM_sim_grouped_exp2}
\end{table}

\begin{table}[hbt!] 
\centering
\begin{tabular}{ccc|cccccc|cc}
\hline\hline
\multicolumn{1}{c}{} &
\multicolumn{2}{c|}{} &
\multicolumn{8}{|c}{{MTuM Performance for Exponential Grouped Data}} \\
\cline{4-11}
\multicolumn{1}{c}{} &
\multicolumn{2}{c|}{} &
\multicolumn{8}{|c}{Sample size, $n$} \\
\cline{4-11}
& $t_{l}$ & $t_{r}$ & 50 & 100 & 250 & 500 & 1000 & $\infty$ & $\infty$ & $\infty$ \\
\hline
& & & & & & & & & & \\[-2.25ex]
MEAN & 0 & 200 &
1.00{\tiny (.004)} & 1.00{\tiny (.003)} & 1.00{\tiny (.002)} &
1.00{\tiny (.002)} & 1.00{\tiny (.001)} & 1 & - & - \\
& 0 & 50 &
1.01{\tiny (.005)} & 1.00{\tiny (.003)} & 1.00{\tiny (.002)} &
1.00{\tiny (.001)} & 1.00{\tiny (.001)} & 1 & - & - \\
& 0 & 100 &
1.00{\tiny (.004)} & 1.00{\tiny (.003)} & 1.00{\tiny (.002)} &
1.00{\tiny (.001)} & 1.00{\tiny (.001)} & 1 & - & - \\
& 0 & 140 &
1.00{\tiny (.004)} & 1.00{\tiny (.004)} & 1.00{\tiny (.003)} &
1.00{\tiny (.002)} & 1.00{\tiny (.001)} & 1 & - & - \\
& 2 & 12 &
1.41{\tiny (.062)} & 1.13{\tiny (.019)} & 1.04{\tiny (.007)} &
1.02{\tiny (.003)} & 1.01{\tiny (.004)} & 1 & - & - \\
\hline\hline
& & & & & & & & & & \\[-2.25ex]
{\sc RE} & 0 & 200 &
0.88{\tiny (.026)} & 0.91{\tiny (.049)} & 0.88{\tiny (.027)} &
0.87{\tiny (.033)} & 0.86{\tiny (.037)} & 0.86 & 0.84 & 0.97 \\
& 0 & 50 &
0.79{\tiny (.043)} & 0.79{\tiny (.044)} & 0.81{\tiny (.045)} &
0.82{\tiny (.019)} & 0.82{\tiny (.028)} & 0.83 & 0.80 & 0.97 \\
& 0 & 100 &
0.92{\tiny (.036)} & 0.94{\tiny (.030)} & 0.94{\tiny (.033)} &
0.94{\tiny (.038)} & 0.94{\tiny (.031)} & 0.95 & 0.92 & 0.97 \\
& 0 & 140 &
1.01{\tiny (.078)} & 0.99{\tiny (.047)} & 1.01{\tiny (.040)} &
0.94{\tiny (.038)} & 1.02{\tiny (.038)} & 1.00 & 0.97 & 0.97 \\
& 2 & 12 &
0.01{\tiny (.002)} & 0.03{\tiny (.007)} & 0.07{\tiny (.005)} &
0.09{\tiny (.004)} & 0.10{\tiny (.006)} & 0.10 & 0.10 & 0.97 \\
\hline\hline
\end{tabular} 
\caption{Finite-sample performance evaluation of MTuM w.r.t. MLE for grouped data from $Exp(\theta=10)$ with group boundaries vector $G_{3} = \left(0:5:50,200\right)$.}
\label{table:MTuM_sim_grouped_exp3}
\end{table}

\begin{table}[hbt!] 
\centering
\begin{tabular}{ccc|cccccc|cc}
\hline\hline
\multicolumn{1}{c}{} &
\multicolumn{2}{c|}{} &
\multicolumn{8}{|c}{{MTuM Performance for Exponential Grouped Data}} \\
\cline{4-11}
\multicolumn{1}{c}{} &
\multicolumn{2}{c|}{} &
\multicolumn{8}{|c}{$n$} \\
\cline{4-11}
& $t_{l}$ & $t_{r}$ & 50 & 100 & 250 & 500 & 1000 & $\infty$ & $\infty$ & $\infty$ \\
\hline
& & & & & & & & & & \\[-2.25ex]
MEAN & 0 & 200 &
1.00{\tiny (.002)} & 1.00{\tiny (.004)} & 1.00{\tiny (.002)} &
1.00{\tiny (.001)} & 1.00{\tiny (.001)} & 1 & - & - \\
& 0 & 50 &
1.01{\tiny (.007)} & 1.00{\tiny (.004)} & 1.00{\tiny (.002)} &
1.00{\tiny (.001)} & 1.00{\tiny (.001)} & 1 & - & - \\
& 0 & 100 &
1.00{\tiny (.006)} & 1.00{\tiny (.003)} & 1.00{\tiny (.002)} &
1.00{\tiny (.002)} & 1.00{\tiny (.001)} & 1 & - & - \\
& 0 & 140 &
1.00{\tiny (.006)} & 1.00{\tiny (.002)} & 1.00{\tiny (.002)} &
1.00{\tiny (.001)} & 1.00{\tiny (.001)} & 1 & - & - \\
& 2 & 12 &
1.16{\tiny (.022)} & 1.06{\tiny (.007)} & 1.02{\tiny (.005)} &
1.01{\tiny (.004)} & 1.01{\tiny (.002)} & 1 & - & - \\
\hline\hline
& & & & & & & & & & \\[-2.25ex]
{\sc RE} & 0 & 200 &
1.00{\tiny (.032)} & 1.03{\tiny (.038)} & 1.00{\tiny (.036)} &
0.99{\tiny (.049)} & 1.01{\tiny (.048)} &  1.00 & 0.92 & 0.92 \\
& 0 & 50 &
0.76{\tiny (.054)} & 0.78{\tiny (.031)} & 0.78{\tiny (.028)} &
0.80{\tiny (.031)} & 0.81{\tiny (.027)} & 0.81 & 0.74 & 0.92 \\
& 0 & 100 &
0.97{\tiny (.064)} & 0.97{\tiny (.038)} & 1.00{\tiny (.034)} &
0.97{\tiny (.053)} & 0.99{\tiny (.026)} & 1.00 & 0.92 & 0.92 \\
& 0 & 140 &
0.99{\tiny (.042)} & 1.00{\tiny (.061)} & 1.01{\tiny (.058} &
1.00{\tiny (.024)} & 1.01{\tiny (.044)} & 1.00 & 0.92 & 0.92 \\
& 2 & 12 &
0.04{\tiny (.010)} & 0.11{\tiny (.020)} & 0.16{\tiny (.010)} &
0.18{\tiny (.007)} & 0.18{\tiny (.008)} & 0.18 & 0.17 & 0.92 \\
\hline\hline
\end{tabular} 
\caption{Finite-sample performance evaluation of MTuM w.r.t. MLE for grouped data from $Exp(\theta=10)$ with group boundaries vector $G_{4} = \left(0:10:100,200\right)$.}
\label{table:MTuM_sim_grouped_exp4}
\end{table}

\begin{table}[hbt!] 
\centering
\begin{tabular}{ccc|cccccc|cc}
\hline\hline
\multicolumn{1}{c}{} &
\multicolumn{2}{c|}{} &
\multicolumn{8}{|c}{{MTuM Performance for Exponential Grouped Data}} \\
\cline{4-11}
\multicolumn{1}{c}{} &
\multicolumn{2}{c|}{} &
\multicolumn{8}{|c}{$n$} \\
\cline{4-11}
& $t_{l}$ & $t_{r}$ & 50 & 100 & 250 & 500 & 1000 & $\infty$ & $\infty$ & $\infty$ \\
\hline
& & & & & & & & & & \\[-2.25ex]
MEAN & 0 & 200 &
0.68{\tiny (.011)} & 0.78{\tiny (.007)} & 0.91{\tiny (.006)} &
0.97{\tiny (.003)} & 0.99{\tiny (.002)} & 1 & - & - \\
& 0 & 50 &
n/a & n/a & n/a &
n/a & n/a & n/a & - & - \\
& 0 & 100 &
0.68{\tiny (.011)} & 0.78{\tiny (.008)} & 0.91{\tiny (.011)} &
0.97{\tiny (.004)} & 0.99{\tiny (.003)} & 1 & - & - \\
& 0 & 140 &
0.68{\tiny (.011)} & 0.78{\tiny (.014)} & 0.92{\tiny (.006)} &
0.97{\tiny (.004)} & 0.99{\tiny (.002)} & 1 & - & - \\
& 2 & 12 &
n/a & n/a & n/a &
n/a & n/a & n/a & - & - \\
\hline\hline
& & & & & & & & & & \\[-2.25ex]
{\sc RE} & 0 & 200 &
0.43{\tiny (.003)} & 0.31{\tiny (.006)} & 0.32{\tiny (.014)} &
0.52{\tiny (.033)} & 0.84{\tiny (.079)} & 1.00 & 0.17 & 0.17 \\
& 0 & 50 &
n/a & n/a & n/a &
n/a & n/a & n/a & - & - \\
& 0 & 100 &
0.43{\tiny (.007)} & 0.30{\tiny (.007)} & 0.32{\tiny (.018)} &
0.53{\tiny (.060)} & 0.84{\tiny (.046)} & 0.97 & 0.17 & 0.17 \\
& 0 & 140 &
0.43{\tiny (.006)} & 0.31{\tiny (.009)} & 0.34{\tiny (.018} &
0.55{\tiny (.030)} & 0.87{\tiny (.058)} & 1.00 & 0.17 & 0.17 \\
& 2 & 12 &
n/a & n/a & n/a &
n/a & n/a & n/a & - & - \\
\hline\hline
\end{tabular} 
\caption{Finite-sample performance evaluation of MTuM w.r.t. MLE for grouped data from $Exp(\theta=10)$ with group boundaries vector $G_{5} = 0:50:200$.}
\label{table:MTuM_sim_grouped_exp5}
\end{table}

From $Exp(\theta = 10)$ and 
for different specific sample sizes 
$n=50, 100, 250, 500, 1000$, 
we generate $1,000$ 
samples, group the sample data with group boundaries
$0=c_{0}<c_{1}<\cdots<c_{m} \leq \infty$ and compute $1,000$ estimated thetas under MTuM with different truncation points for grouped data as $\widehat{\theta}_{1},\ldots,\widehat{\theta}_{1000}$.
Set
$
\displaystyle 
\bar{\widehat{\theta}}
= \left( \sum_{i=1}^{1000}\hat{\theta}_{i} \right)/1000$. 
We repeat this process $10$ times getting $\bar{\hat{\theta}}_{1},\ldots ,\bar{\hat{\theta}}_{10}$. Then we compute mean $\hat{\hat{\theta}}$ and standard deviation $se(\bar{\hat{\theta}})$ of $\bar{\hat{\theta}}_{1},\ldots ,\bar{\hat{\theta}}_{10}$ and finally $\frac{\hat{\hat{\theta}}}{\theta}$ and $\frac{se(\bar{\hat{\theta}})}{\theta}$ are reported on the table. 
Similarly, finite-sample relative efficiency (RE) of MTuM w.r.t. grouped MLE are computed as $RE_{1},\ldots ,RE_{10}$ and their mean, standard deviations are reported for different vectors of group boundaries. 

The following vectors of group boundaries were considered:
\begin{align*}
G_{1} &:= \left(0:1:100,200\right) \, 
G_{2} := 0:1:200, \ 
G_{3} := \left(0:5:50,200\right), \\
G_{4} &:= \left(0:10:100,200\right), 
\ \mbox{and} \
G_{5} := 0:50:200.
\end{align*}

The outcomes of the simulations are documented in Tables 
\ref{table:MTuM_sim_grouped_exp1}-\ref{table:MTuM_sim_grouped_exp5}. 
The entries are mean values 
(with standard errors in parentheses).
In all tables, 
the last three columns (with $\infty$)
represent analytic results, 
not from simulation.
The third last column is for the asymptotic 
relative efficiency of MTuM w.r.t. grouped MLE. 
Similarly, the second last column is for the 
asymptotic relative efficiency of MTuM w.r.t
un-grouped MLE and the very last column 
represents the asymptotic relative efficiency
of grouped MLE w.r.t. un-grouped MLE. 
If both the truncation points are in
the same interval,
say $t_{l}, t_{r} \in [c_{j-1}, c_{j}]$
then we have $\widehat{\mu} 
= \mu 
= \frac{t_{l} + t_{r}}{2}$. 
Therefore, the parameter $\theta =10$ 
to be estimated disappeared and hence
the four rows on 
Table \ref{table:MTuM_sim_grouped_exp5} 
are reported as $n/a$.
As we move in sequence from Table 
\ref{table:MTuM_sim_grouped_exp1}
to Table \ref{table:MTuM_sim_grouped_exp5},
it becomes noticeable that the convergence of 
the ratio of the estimated $\theta$ with the true $\theta$, 
i.e., $\widehat{\theta}/\theta$,
\label{mis:Mean2}
approaches the true asymptotic value of 1 at 
a more gradual pace/slower.
Additionally, 
both our intuition and the data presented 
in the tables suggest that when 
there is a wider gap between the thresholds
(namely, $t$ and $T$), 
the estimators tend to approach 
the true values at a slower rate.

\section{Concluding Remarks}
\label{sec:Conclusion}

In this scholarly work,
we have purposed a novel 
{\em \underline{M}ethod of 
\underline{T}r\underline{u}ncated 
\underline{M}oments} (MTuM)
estimator to estimate the tail index 
from grouped Pareto loss severity data, 
as a robust alternative to MLE. 
Theoretical justifications regarding the
designed estimators' existence and asymptotic
normality are established. 
The finite sample performance, 
for various sample sizes and different
group boundaries vectors has been investigated
in detail via simulation study.

Regarding future directions, 
this paper focused mainly on 
estimating the mean parameter 
of an exponential distribution
(equivalently estimating tail index 
of a single parameter Pareto distribution)
given a grouped sample data, 
so the purposed methodology could be 
extended to more general situations and models.
However,
and specially for multi-parameter distributions, 
it is very complicated to investigate the nature 
of the function $g_{tT}(\mbox{Parameters})$
given in Eq. \eqref{eqn:gTt1} and 
Conjecture \ref{conj:gtTGrouped},
if not intractable. 
Asymptotic inferential justification of 
the designed MTuM methodology for multi-parameter
distributions is equally difficult. 
In this regard,
a potential future direction is to 
think algorithmically 
\citep[i.e., designing simulation based estimators for complex models,][]{MR3941244}
rather than establishing inferential justification,
\citet[][p. xvi]{MR3523956}. 
Furthermore, 
it remains to be assessed how this novel MTuM estimator 
performs across various practical risk analysis scenarios.

{\baselineskip 5.50mm
\renewcommand{\bibname}{References}
\bibliography{gMTuM_ArXiv}

\begin{thebibliography}{}

\bibitem[\protect\astroncite{Aigner and Goldberger}{1970}]{ag70}
Aigner, D.~J. and Goldberger, A.~S. (1970).
\newblock Estimation of {P}areto's law from grouped observations.
\newblock {\em Journal of the American Statistical Association},
  65(330):712--723.

\bibitem[\protect\astroncite{Beran}{1977a}]{MR0448700}
Beran, R. (1977a).
\newblock Minimum {H}ellinger distance estimates for parametric models.
\newblock {\em The Annals of Statistics}, 5(3):445--463.

\bibitem[\protect\astroncite{Beran}{1977b}]{MR0448699}
Beran, R. (1977b).
\newblock Robust location estimates.
\newblock {\em The Annals of Statistics}, 5(3):431--444.

\bibitem[\protect\astroncite{Brazauskas et~al.}{2009}]{MR2497558}
Brazauskas, V., Jones, B.~L., and Zitikis, R. (2009).
\newblock Robust fitting of claim severity distributions and the method of
  trimmed moments.
\newblock {\em Journal of Statistical Planning and Inference},
  139(6):2028--2043.

\bibitem[\protect\astroncite{Chernoff et~al.}{1967}]{MR0203874}
Chernoff, H., Gastwirth, J.~L., and Johns, Jr., M.~V. (1967).
\newblock Asymptotic distribution of linear combinations of functions of order
  statistics with applications to estimation.
\newblock {\em Annals of Mathematical Statistics}, 38(1):52--72.

\bibitem[\protect\astroncite{Efron and Hastie}{2016}]{MR3523956}
Efron, B. and Hastie, T. (2016).
\newblock {\em Computer Age Statistical Inference: Algorithms, Evidence, and
  Data Science}.
\newblock Cambridge University Press, New York.

\bibitem[\protect\astroncite{Gatti and W\"{u}thrich}{2023}]{gw23}
Gatti, S. and W\"{u}thrich, M.~V. (2023).
\newblock Modeling lower-truncated and right-censored insurance claims with an
  extension of the {MBBEFD} class.
\newblock {\em ArXiv:2310.11471}, pages 1--28.

\bibitem[\protect\astroncite{Guerrier et~al.}{2019}]{MR3941244}
Guerrier, S., Dupuis-Lozeron, E., Ma, Y., and Victoria-Feser, M.-P. (2019).
\newblock Simulation-based bias correction methods for complex models.
\newblock {\em Journal of the American Statistical Association},
  114(525):146--157.

\bibitem[\protect\astroncite{Hongqi and Lixin}{2002}]{MR1900948}
Hongqi, X. and Lixin, S. (2002).
\newblock Asymptotic properties of {MLE} for {W}eibull distribution with
  grouped data.
\newblock {\em Journal of Systems Science and Complexity}, 15(2):176--186.

\bibitem[\protect\astroncite{Kleiber and Kotz}{2003}]{MR1994050}
Kleiber, C. and Kotz, S. (2003).
\newblock {\em Statistical Size Distributions in Economics and Actuarial
  Sciences}.
\newblock John Wiley \& Sons, Hoboken, NJ.

\bibitem[\protect\astroncite{Klugman et~al.}{2019}]{MR3890025}
Klugman, S.~A., Panjer, H.~H., and Willmot, G.~E. (2019).
\newblock {\em Loss Models: From Data to Decisions}.
\newblock John Wiley \& Sons, Hoboken, NJ, fifth edition.

\bibitem[\protect\astroncite{Lin and He}{2006}]{MR2277743}
Lin, N. and He, X. (2006).
\newblock Robust and efficient estimation under data grouping.
\newblock {\em Biometrika}, 93(1):99--112.

\bibitem[\protect\astroncite{Poudyal}{2021a}]{MR4263275}
Poudyal, C. (2021a).
\newblock Robust estimation of loss models for lognormal insurance payment
  severity data.
\newblock {\em ASTIN Bulletin. The Journal of the International Actuarial
  Association}, 51(2):475--507.

\bibitem[\protect\astroncite{Poudyal}{2021b}]{MR4192140}
Poudyal, C. (2021b).
\newblock Truncated, censored, and actuarial payment-type moments for robust
  fitting of a single-parameter {P}areto distribution.
\newblock {\em Journal of Computational and Applied Mathematics}, 388:113310,
  18.

\bibitem[\protect\astroncite{Poudyal and Brazauskas}{2022}]{pb22}
Poudyal, C. and Brazauskas, V. (2022).
\newblock Robust estimation of loss models for truncated and censored severity
  data.
\newblock {\em Variance -- The scientific journal of the Casualty Actuarial
  Society}, 15(2):1--20.

\bibitem[\protect\astroncite{Poudyal and Brazauskas}{2023}]{pb23}
Poudyal, C. and Brazauskas, V. (2023).
\newblock Finite-sample performance of the $t$- and $w$-estimators for the
  pareto tail index under data truncation and censoring.
\newblock {\em Journal of Statistical Computation and Simulation},
  93(10):1601--1621.

\bibitem[\protect\astroncite{Poudyal et~al.}{2023}]{pzb23}
Poudyal, C., Zhao, Q., and Brazauskas, V. (2023).
\newblock Method of winsorized moments for robust fitting of truncated and
  censored lognormal distributions.
\newblock {\em North American Actuarial Journal}, pages 1--25.

\bibitem[\protect\astroncite{Schader and Schmid}{1986}]{MR0909389}
Schader, M. and Schmid, F. (1986).
\newblock Optimal grouping of data from some skew distributions.
\newblock {\em Computational Statistics Quarterly}, 3(3):151--159.

\bibitem[\protect\astroncite{Serfling}{1980}]{MR595165}
Serfling, R.~J. (1980).
\newblock {\em Approximation Theorems of Mathematical Statistics}.
\newblock John Wiley \& Sons, New York.

\bibitem[\protect\astroncite{Tukey}{1960}]{MR0120720}
Tukey, J.~W. (1960).
\newblock A survey of sampling from contaminated distributions.
\newblock In {\em Contributions to Probability and Statistics}, pages 448--485.
  Stanford University Press, Stanford, CA.

\bibitem[\protect\astroncite{Victoria-Feser and Ronchetti}{1997}]{MR1436121}
Victoria-Feser, M.-P. and Ronchetti, E. (1997).
\newblock Robust estimation for grouped data.
\newblock {\em Journal of the American Statistical Association},
  92(437):333--340.

\bibitem[\protect\astroncite{Zhao et~al.}{2018}]{MR3758788}
Zhao, Q., Brazauskas, V., and Ghorai, J. (2018).
\newblock Robust and efficient fitting of severity models and the method of
  {W}insorized moments.
\newblock {\em ASTIN Bulletin}, 48(1):275--309.

\end{thebibliography}
\thispagestyle{plain}
\pagestyle{plain} \thispagestyle{plain}
}

\end{document}